\documentclass[preprint2]{aastex}

\usepackage{natbib}

\usepackage{graphicx}
\usepackage{multirow}
\usepackage{txfonts}

\shorttitle{Mass Estimation of H~1743-322}
\shortauthors{A. A. Molla, S. K. Chakrabarti, D. Debnath, \& S. Mondal}

\begin{document}

\title{Estimation of Mass of Compact Object in H 1743-322 from 2010 and 2011 Outbursts using TCAF Solution and Spectral Index - QPO Frequency Correlation}

\author{ Aslam Ali Molla\altaffilmark{1}, Sandip K. Chakrabarti\altaffilmark{2,1}, Dipak Debnath\altaffilmark{1}, Santanu Mondal\altaffilmark{3,1}}
\altaffiltext{1}{Indian Center for Space Physics, 43 Chalantika, Garia St. Rd., Kolkata, 700084, India.}
\altaffiltext{2}{S. N. Bose National Centre for Basic Sciences, Salt Lake, Kolkata, 700098, India.}
\altaffiltext{3}{Instituto de F\'isica y Astronom\'ia, Facultad de Ciencias, Universidad de Valpara\'iso, Gran Bretana N 1111, Playa Ancha, Valparaíso, Chile}

\email{aslam@csp.res.in; chakraba@bose.res.in; dipak@csp.res.in; santanu@csp.res.in}


\begin{abstract}

The well known black hole candidate H~1743-322 exhibited temporal and spectral variabilities during several outbursts. 
Daily variation of the accretion rates and the flow geometry change on a daily basis during each of the outbursts 
could be understood very well using the recent implementation of two component advective flow (TCAF) solution of 
the viscous transonic flow equations as an additive table model in XSPEC. This has dramatically improved our understanding about 
the accretion flow dynamics.  Most interestingly, the solution allows to treat mass of the black hole 
candidate as a free parameter and there mass could be estimated from spectral fits. 
In this paper, we fit the data of two successive outbursts of H~1743-322 in 2010 and 
2011 and studied evolutions of accretion flow parameters, such as, two component (Keplerian and sub-Keplerian) accretion 
rates, shock location (i.e., size of the Compton cloud), etc. 
We assume that the model Normalization remains the same accross the states in both these outbursts. 
We use this to estimate mass of the black hole and found that 
it comes out in the range of $9.25-12.86 M_\odot$. For the sake of comparison, 
we also estimated mass using Photon index vs. QPO frequency correlation method which turns out to 
be $11.65 \pm 0.67 M_\odot$ using GRO J1655-40 as reference source. Combining these two estimates, the most probable mass of the 
compact object becomes $11.21^{+1.65}_{-1.96} M_\odot$. 

\end{abstract}

\keywords{X-Rays:binaries -- stars individual: (H 1743-322) -- stars:black holes -- accretion, accretion disks -- shock waves -- radiation:dynamics}

\section{Introduction}

In the last two decades, specially after the launch of {\it Rossi X-Ray Timing Explorer (RXTE)}, our understanding about 
the accretion process around a black hole candidate (BHC) is significantly improved. They are only detectable when matter 
from a companion, be it from the Roche lobe or from its winds accretes on them. Radiations are emitted in this process 
allowing them to infer on the accretion process. It is even better, when the rate of accretion changes rapidly, 
since the object can then go through various spectral states in quick successions and the timing properties also 
give out information about the companion. Normally, an outbursting BHC shows four different spectral states, 
namely, hard (HS), hard-intermediate (HIMS), soft-intermediate (SIMS), and soft (SS) (Belloni et al. 2005; 
McClintock \& Remillard 2006; Nandi et al. 2012; Debnath et al. 2013). Low frequency Quasi Periodic Oscillations (QPOs) 
in some of these states are also observed (e.g., Remillard \& McClintock 2006). Evolution of temporal and spectral 
properties of several BHCs have been studied by many workers during their outbursts (e.g., Belloni et al., 2005; 
McClintock \& Remillard, 2006; Nandi et al., 2012) and it has been noted by several authors that different spectral 
states are related to different branches of hardness intensity diagram (HID; Belloni et al. 2005; Debnath et al. 2008) 
or accretion rate ratio intensity diagram (ARRID; Jana et al. 2016).

It is well known that in order to explain majority of the black hole spectra one needs two types of components: 
one is a multi-color blackbody component radiated by the standard Keplerian disk (Shakura \& Sunyaev 1973) and 
the other is a power-law (PL) component, originated from a so-called ``Compton cloud" (Sunyaev \& Titarchuk 1980, 1985). 
There are several models which describe the nature and origin of this ``Compton cloud". These range from a magnetic 
corona (Galeev et al. 1979) to a hot gas corona over the disk (Haardt \& Maraschi 1993; Zdziarski et al. 2003). 
Chakrabarti \& Titarchuk (1995; hereafter CT95), even before the launch of RXTE, considered the well known component solution 
of transonic flows (see, Chakrabarti 1990; Chakrabarti 1996), namely, the CENtrifugal Pressure supported BOundary Layer (CENBOL) 
to represent the ``Compton cloud" in their Two Component Advective Flow (TCAF) solution. 
From the observational evidence it is now established that, both the components are dynamic and the low angular 
momentum halo component (which does not require viscosity to accrete) is moving faster than the more viscous 
Keplerian disk (Soria et al. 2001; Smith et al. 2002; Wu et al. 2002; Cambier \& Smith 2013; Tomsick et al. 2014). 
In the early stage of an outburst, the gravitational energy is mostly stored in the advective component as thermal
energy and as the day progresses, this energy is released gradually by inverse Comptonization since the soft (seed) photon flux 
also rises. In the declining phase the process is reversed, though the time scale varies.
Thus the hot post-shock flow in HS becomes cooler progressively in HIMS, SIMS and SS. Whether or not all the states 
would be visited will depend on the how long the supply of viscous accretion creating the Keplerian disk lasts.
In a transonic flow with low enough viscosity, the centrifugal pressure slows down the flow and a standing or oscillating 
shock would be formed depending on whether the Rankine-Hugoniot conditions are satisfied (Chakrabarti 1989, 1996) or not. 
This centrifugal barrier region is puffed up and acts as the ``Compton cloud".
Recently, self-consistency and stability check (Giri \& Chakrabarti 2013; Mondal \& Chakrabarti 2013)
of the transonic solution have been made which support from both the hydrodynamic and radiative points 
of view that an advective flow will desegregate into a TCAF solution (CT95) when viscous stress near the equatorial 
plane is very high.

Successful fits of data from several black hole candidates have been recently obtained and the general nature 
of the accretion flow dynamics has been understood (Debnath, Chakrabarti \& Mondal 2014; 
Mondal, Debnath \& Chakrabarti 2014; Debnath, Mondal \& Chakrabarti 2015a; 
Debnath, Molla, Chakrabarti \& Mondal 2015b; Jana et al. 2016; Molla et al. 2016;
Chatterjee et al. 2016; Mondal, Chakrabarti \& Debnath 2016). Physical flow parameters such as the two types 
of accretion rates (disk and halo), shock location (size of the Compton cloud), shock strength and the mass 
of the black hole were obtained after the inclusion of TCAF solution (CT95, Chakrabarti 1997) as an additive table model in HEASARC's 
spectral analysis software package XSPEC. 

The black hole candidate H~1743-322 is very active as it displayed several X-ray outbursts since its first 
detection in 1977 August-September by the {\it Ariel V} all sky monitor (Kaluzeinski \& Holt 1997) and 
{\it HEAO I} satellite (Doxsey et al. 1977). After its first detection, it went to the 
quiescent state and remained that way for a long time till 1984. A couple of X-ray activities were reported 
by {\it EXOSAT} observations in 1984 (Raynolds 1999) and by TTM/COMIS instruments on board {\it Mir-Kvant} 
in 1996 (Emelyanov et al., 2000). During the scanning of the galactic center on March 21, 2003, {\it INTEGRAL} 
detected a bright source named IGR~J17464-3213 (Revnivtsev et al. 2003) which turned out to be H~1743-322 
(Markwardt \& Swank 2003). Since then, it exhibited several X-ray activities with regular intervals of 
one or two years. This active low mass X-ray binary source is located at the sky location of 
R.A. $= 17^h46^m15^s{.61}$, Dec $= -32^\circ 14'00''.6$ (Gursky et al., 1978). Mass of this BHC has 
not been dynamically measured yet, although Petri (2008), with their high frequency QPO model estimated 
that its mass could be between 9$M_\odot$ and 13$M_\odot$. From the correlation between spectral and timing properties
of black holes, Shaposhnikov and Titarchuk (2009; hereafter ST09) calculated mass of this black hole candidate to be $13.3\pm3.2$ $M_\odot$ 
using 2003 data of RXTE PCA by taking GX~339-4 as a reference source of known mass. The inclination angle was reported to be 
about $\theta= 75^\circ\pm 3^\circ$ (McClintock et al. 2009). The source is located at a distance of $8.5\pm0.8$ kpc with 
an angle of inclination  and spin of $-0.3 <$ a $< 0.7$, with a 90\% confidence (Steiner et al. 2012).  

It is well known that timing and spectral properties of a BH are tightly correlated. Titarchuk \& Fiorito (2004), 
introduces a useful tool to determine mass of stellar massive black hole as well as super massive black hole by the 
correlation between power-law (PL) photon indices and characteristic frequencies of the observed quasi-periodic oscillation. 
Shaposhnikov \& Titarchuk (2007; hereafter ST07) employed this idea to estimate the mass of Cyg~X-1 and obtained the mass to be
$M_{CygX-1} = 8.7 \pm 0.8 M_\odot$. This method was also used by Fiorito \& Titarchuk (2004), Dewangan, Titarchuk \& Griffiths (2006) 
and Strohmayer et al. (2007), Titarchuk \& Seifina (2016a,b) to estimate masses of a number of ultra-luminous X-ray sources. 
This idea was applied successfully to calculate mass of an AGN Mrk 766 by Giacche et al. (2014) to obtain 
$M_{BH} = 1.26^{+1.00}_{-0.77} \times 10^6 M_\odot$. Papadakis et al. (2009) and Sobolewska \& Papadakis (2009) 
have applied this method to estimate the mass of 14 AGNs. 
Using a simplified accretion disk model of non-interacting particles, Petri (2008) also presented a way to estimate 
the mass and spin of BHCs. However, this method required another reference source to estimate the mass. Furthermore, 
the error-bar obtained by ST09 on estimated mass of H~1743-322 is also quite large. Most importantly, since 2003 outburst 
of H~1743-322 is quite anomalously bright with other effects such as activities in radio, which means the jets could also 
contribute to X-rays, it was essential to apply this method to a more `normal' outburst event which occured more recently. 
This motivated us to estimate the mass once more using more recent outbursts.

In, 2010 and 2011, H~1743-322 exhibited two X-ray flaring activities (Yamaoka et al. 2010; 
Kuulkers et al. 2011) with the same type of state transition characteristics (Debnath et al. 2013) as 
reported in other outbursting sources (Debnath et al. 2008, 2015a,b; Jana et al. 2016, Chatterjee et al. 2016  
and references therein). Both the outbursts were covered by {\it RXTE} almost on a daily basis for around 
two months during the outbursts. Study of gross aspects of the temporal and spectral properties during 
these two outbursts using {\it RXTE PCA} data with combined disk black body (DBB) and power law (PL) 
model has already been made by Debnath et al. (2013). Study of spectral property using {\it RXTE PCA} 
data during its 2010 outburst with {\it TCAF} solution has also been done quite recently by
Mondal et al. (2014), where TCAF model fitted normalization (N) was allowed to vary for the best fit. 
In the present paper, we restrict the normalization N within a narrow range
to estimate the mass of the central object independently from two successive outbursts (2010 
and 2011) of H~1743-322, in order to investigate if there are still consistent fits from the TCAF solution.
This would indicate that no free parameter other than those five (mass, two accretion rates, 
shock location and shock compression ratio) required by TCAF is necessary to explain the entire outburst.  
N would otherwise be expected to remain constant as it is an intrinsic property of the system 
from one outburst to another, though it could have an 
error-bar consistent with the error-bars of other parameters on which it depends, such as the mass, 
distance and the inclination angle. 
Systematic deviation of N from a constant value could be due to several effects such as the 
precession in the disk occurs which changes the projected surface area of emission. Significant fluctuations in N could be interpreted 
as contributions of X-rays from jets or outflows which have not been included in this version of TCAF model {\it fits} file.

In Molla et al. (2016), Jana et al. (2016) and Chatterjee (2016), we successfully applied this constant normalization method 
to estimate masses of three MAXI transient compact objects, such as, MAXI~J1659-152, MAXI~1836-194, and MAXI~J1543-564 respectively. 
In Molla et al. (2016), we also verified estimated mass of MAXI~J1659-152 from that obtained by us using propagating 
oscillatory shock (POS) model (Chakrabarti et al. 2005, 2008; Debnath et al. 2010, 2013). The successful interpretation of masses 
of these MAXI objects also motivated us to estimate mass of the well known transient BHC H~1743-322 using the same method.

The organization of this {\it paper} is the following: in the next Section, we discuss observation and data 
analysis procedures. In \S 3, we present results of TCAF model fitted spectral analysis. 
From this we estimate the mass of the object. We also compare with the mass obtained from ST09. 
In the final Section, we present a brief discussion and draw concluding remarks.

\section{Observation and Data Analysis}

We take RXTE/PCA science data from NASA data archive for analysis. We choose the data produced in 2010 and 2011 by RXTE 
as in this period H~1743-322 show spectral and timing property very clearly almost on a daily basis. We analyze the data 
of $26$ observational IDs for 2010 outburst starting from 2010 August 9 (MJD=55417) to 2010 September 30 (MJD=55469) and 
the data of $27$ observational IDs for 2011 outburst starting from 2011 April 12 (MJD=55663) to 2011 May 19 (MJD=55700). 
In our data analysis we exclude the data taken for elevation angles less than $10^\circ$, for offset greater than $0.02^\circ$ 
and those acquired during the South Atlantic Anomaly (SAA) passage. We use HEASARC's software package HeaSoft version 
HEADAS 6.15 and XSPEC version 12.8 to carry out our timing and spectral data analysis procedure.

\subsection{Timing Analysis}

In order to observe timing property of RXTE/PCA data we use ``binned mode" data which was available for 0-35 channels only, with 
a maximum time resolution of $125~\mu$s. For the timing analysis of PCA data we restrict ourselves in the energy range of $\sim 2-15$~keV.
After extracting the light curve, Power Density Spectrum (PDS) is generated by the standard FTOOLS task ``powspec" with a suitable 
normalization. We find that the power beyond $50$~Hz frequency is insignificant so we rebinned the data to have $0.01$s time resolution 
to obtain the Nyquest frequency of 50 Hz and also to get the $QPO$ frequency. In Figure 1, we show the PDS with a QPO having primary 
frequency of $1.008 \pm 0.003$~Hz and the total exposure time of data is $3379$s.

\subsection{Spectral Analysis}

In order to carry out the spectral analysis of the PCA data, we use ``standard2" mode Science Data (FS4a*.gz)
in the energy range 2.5-25. keV with a time resolution of 16 sec. We extract spectra from all the layers of PCU2 for 128 channels 
(without any binning/grouping the channels). We generate source spectrum using FTOOLS task ``SEEXTRCT" with 16 sec time bin
from ``standard2" science even mode data. In order to produce background {\it fits} file we use FTOOLS task ``runpcabackest" with
the standard FILTER file provided with the package. To generate background source spectrum we again use FTOOLS task ``SEEXTRCT"
with 16 sec time bin from background fits file. The ``pcarsp" task is used to produce the response file with appropriate detector information.
To fit the spectra we use TCAF based model {\it fits} file. In order to achieve the best fit, we use Gaussian line of peak energies 
around $6.5$~keV and around $3.8$~keV for both the outbursts. Spectral data is fitted with model composition
wabs$\times$(TCAF+Gaussian+Gaussian). The hydrogen column density (N$_{H}$) obtained for this source was 1.6$\times$10$^{21}$~atoms~cm$^{-2}$ 
(Capitanio et al., 2009) from Swift/XRT observations, which was lower energy range than that for RXTE/PCA and therefore we keep 
it fixed for both 2010 and 2011 outburst for absorption model {\it wabs}. We also use a fixed $0.5$\% systematic instrumental error.
We use ``err" command to find out 90\% confidence error values in model fitted parameters. During a spectral fit, using the TCAF 
based fits file, we supply five model input parameters:
$i)$ black hole mass ($M_{BH}$) in solar mass ($M_\odot$) unit, $ii)$ sub-Keplerian rate ($\dot{m_h}$ in $\dot{M}_{Edd}$), $iii)$ Keplerian rate 
($\dot{m_d}$ in Eddington rate $\dot{M}_{Edd}$), $iv)$ location of the shock ($X_s$ in Schwarzschild radius $r_g$=$2GM/c^2$), $v)$ 
compression ratio ($R$) of the shock. To fit a black hole spectrum with the TCAF in XSPEC, we generated model {\it fits} file 
(here {\it TCAF0.3.fits} was used). For this, we use theoretical spectra generating software and vary our basic five 
input parameters in suitably generalized CT95 code and then include it in XSPEC as a local additive model. 
During the analysis of both the outbursts we keep all the parameters free, except with the restriction 
that the N which comes of the fit must be in a narrow range, i.e., a constant with a small enough error bar.
In order to extract photon index from spectral data, we refit entire 2010 and 2011 outburst data sets using 
CompTB model (Farinelli et al. 2008) with spectral fitting model composition as wabs$\times$(CompTB+Gaussian).

\section{Results} 

Here we present results based on spectral analysis using TCAF {\it fits} file. Classification of different spectral states and their 
transitions during both 2010 and 2011 outbursts of H~1743-322 are done on the basis of the values of flow parameters extracted from 
TCAF {\it fits}, and values of the ``QPO frequency". These classifications are consistent with Debnath et al. (2013).

\subsection{Spectral Data Fitted by TCAF model} 

In Figs. 2 \& 4, we show variation of background subtracted {\it RXTE} PCA count rate in the energy range of 
2-25 keV (0-58 channels) and QPO frequency along with model fitted parameters during the 2010 and 2011 outbursts. 
From TCAF {\it fits} we extract physical parameters such as the disk rate ($\dot{m_d}$), the halo rate ($\dot{m_h}$), 
the location of shock ($X_s$) representing the size of the Compton cloud, and the compression ratio ($R$). Variations of 
these derived parameters are shown in Figs. 2(b-e) \& 4(b-e). During both the outbursts, we fit 2.5-25 keV PCA spectral 
data with TCAF solution by leaving all the parameters free. From the fit, we get the variation of model normalization and mass. 
By restricting the model normalization in a narrow range allowed by our criteria of acceptable fit ($\chi^2_{red}<2.0$), 
we obtain the mass range to be $9.51 M_\odot $ - $12.42 M_\odot$ for 2010 outburst and $9.43 M_\odot$ - $12.71 M_\odot$ 
for 2011 outburst. In Figs. 3 \& 5, we show variation of model normalization and derived mass of the BHC H~1743-322 using these 
respective data. Variation of normalization and mass are shown in Figs. 3 (a) \& (b) for the 2010 outburst and in Figs. 5 (a) \& (b) 
for the 2011 outburst. As a cross check, we repeat the fitting procedure by keeping model normalization fixed to a value of 
15.55, which is the averaged value of model normalizations obtained from the fits above and we obtain 
a variation of mass to be in the range of $9.63 M_\odot$ - $12.34 M_\odot$ for 2010 outburst and $9.25 M_\odot$ - $12.86 M_\odot$ 
for 2011 outburst. The plots are shown in Figs. 3 (c) and 5 (c). Depending on the extracted accretion 
flow parameters and nature of QPOs (shape, frequency, Q-values, rms), we identify four spectral states, namely, 
HS, HIMS, SIMS, and SS. We observe that the transitions (marked with vertical dashed lines in Fig. 2) occur almost 
on the same date for 2010 outburst (as reported in Mondal et al. 2014, 2015) where the same data was fitted by TCAF 
keeping mass as a constant parameter (11.4 $M_\odot$) and using free normalization constant during the whole outburst. 
TCAF model fitted unfolded spectra are shown in Fig. 6 from HS and SS. The possible reason of
double Gaussian is discussed later. 
In Debnath et al. (2013) spectral classification was made on the basis of the variation of DBB and PL model 
fitted fluxes and properties of QPOs. During the 2011 outburst, the spectral transition dates do not exactly 
match with the transition dates as reported in Debnath et al. (2013). In Appendix Table I, II, III \& IV these fitted 
and derived parameters are presented with parameter error values.
The evolution of the spectral states, observed QPO frequencies and variation of model normalization 
values and mass of the black hole obtained from each observation during 2010 \& 2011 outbursts are also given in the tables.
As a cross check, we find that if we freeze all the model parameters
to the values obtained from their best fit and try to see the variation of $\chi^2_{red}$ by only
changing its mass, then we see that value of $\chi^2_{red} \geq 2$ when the mass of the black hole
is chosen outside the range of $9 M_\odot - 13 M_\odot$. 
All these results are shown in Figs. 7 \& 8. We also measured mass of the source using ST07 photon index - QPO Freq. 
correlation method (see, Fig. 9), which is initially refereed as a BH scaling method (Fiorito \& Titarchuk 2004; ST07). 
The estimated mass (=$11.65 \pm 0.67 M_\odot$) from this method is well within TCAF model fitted mass range. 


\subsection{Spectral evolution during 2010 and 2011 Outburst}

Detailed study of temporal and spectral evolution of this source during 2010 \& 2011 outburst with DBB-PL model (Debnath et al., 2013)
and spectral evolution of 2010 outburst with TCAF solution (Mondal et al., 2014) has already been done. The analysis indicates transition 
of spectral states and a clearer picture of accretion flow dynamics emerged. However, since the mass of the object was not known dynamically,
a suitable value of the mass was chosen and the normalization factor was allowed to vary for an acceptable fit. In the present paper,
on the other hand, we concentrate on the determination of mass by restricting the normalization to be in the narrowest possible range, as allowed by 
acceptable range of $\chi^2_{red}<2.0$. This is because the normalization in TCAF is a function of the mass, distance, and the inclination
angle only, after the data is corrected  due to the instrumental response and the absorption. This gives a narrow range on the mass.
After obtaining the fit for the 2010 outburst, we fit the 2011 outburst also for the same normalization factor range. This is to find out
if the same normalization factor also yields a similar mass as obtained in 2010 outburst.

\noindent{\it (i) Hard-State in the Rising phase:} In hard state, the spectrum is dominated by hard photons. The source
remains in this state for the initial $3$ days of the outburst (from MJD = 55417.29 to 55419.11) in 2010, and for $6$ days of the outburst
(from MJD=55663.68 to 55668.48) in 2011 with very low disk rate and comparatively high halo rate. Changes in the shock location, compression ratio 
as well as the QPO frequency are slow (see, Fig. 2 and 4). We define third day of observation (MJD =55420.3) for 2010 outburst and fifth 
day of observation (MJD=55667) for 2011 outburst as the transition day from HS to HIMS since on the transition day value of halo rate for both the 
outbursts has attained its maximum value in the rising phase and after this day disk rate increases very rapidly
($0.011 $ to $0.039$ for 2010 and $0.013$ to $0.075$ for 2011), and the location of the shock as well as the compression ratio decreases very 
rapidly for both the outbursts. QPO frequency changes very rapidly for both the outbursts ($1.045$Hz to $1.174$Hz for 2010 and $0.885$Hz to 
$1.364$Hz for 2011 outburst).  

\noindent{\it (ii) Hard-Intermediate State in Rising phase:} The source remains in this state for the next four days in 2010 and
for the next six days in 2011 until the next transition occurred on MJD =55424.1 during 2010 and MJD=55672.9 during 2011 outburst.
During this period the spectrum becomes somewhat softer which is reflected in the increased value of Keplerian rate ($0.039$ to $1.008$ 
$\dot{M}_{Edd}$ for 2010 and $0.075$ to $0.286$ $\dot{M}_{Edd}$ for 2011). During this phase, the shock moves closer to the black hole and   
became weaker ($R\sim 1.981$ to $1.075$ for 2010 and $R\sim 1.754$ to $1.076$ for 2011). This is because the post-shock region which acts as 
the Compton cloud is cooled down by increasing number of seed (soft) photons. QPO frequency  increases monotonically and reaches 
to its maximum value indicating an end of this phase.

\noindent{\it (iii) Soft-Intermediate State in Rising phase:} The spectrum makes a transition from HIMS to SIMS when the resonance 
nature of QPO frequency (Molteni et al. 1996, hereafter MSC96; Chakrabarti et al. 2015, hereafter CMD15) breaks down for both the outbursts 
and the value of disk rate further increased by causing the spectrum more soft. The source remains in this state 
for a very short period about $1$ day in 2010 and about $3.5$ days in 2011. In this state, the spectrum becomes softer 
causing an increase in the Keplerian rate and decrease in the sub-Keplerian rate although on MJD = 55674.3 halo rate is very 
high (0.794 $M_{Edd}$) it may be due to high photon count on that day (see top panel of Fig. 4).
Increase in the disk rate cools down the post shock region very rapidly, so the 
shock location and its strength decreases very sharply. In this state LFQPO occurs sporadically as the shock itself may not form 
and the centrifugal barrier oscillates (Debnath et al, 2013). This is the common feature of soft intermediate state. 

\noindent{\it (iv) Soft State:} After the completion of the soft intermediate state, the source enters into the 
soft state with a high Keplerian rate and very low sub-Keplerian rate. The shock 
location and the shock strength remain almost constant. The source remains in this state for $24$ days 
(MJD= $55425.2$ to $55448.8$) in 2010 outburst and for about $9.5$ days (MJD= $55678.1$ to $55687.6$) in 2011 outburst. No sign of QPOs in this period
in both the outbursts as the cooling time scale is very short and the resonance condition cannot be satisfied.  

\noindent{\it (v) Soft-Intermediate State in Declining phase:} The source enters in this state with when sporadic QPOs start and remains
in this state for almost $\sim$ 4 days (MJD = $55450.3$ to $55453.7$) and for $2.5$ days (MJD=$55687.6$ to MJD=$55690.1$) respectively for 2010
and 2011 outbursts. In 2010 outburst, it has shown a sporadic QPO with the peak frequency of $2.489 \pm 0.018$Hz whereas in 2011 outburst it has shown only one
sporadic QPO with a frequency of $2.215 \pm 0.021$Hz. In this state, the total PCA rate as well as the total flow rate ($\dot{m}_d + \dot{m}_h$) are nearly constant for both the outbursts. After this state, on  MJD = $55455.4$ and on MJD = $55690.1$, a QPO was observed with a much higher frequency of $6.417 \pm 0.252$ Hz (2010 outburst) and $2.936 \pm 0.024$ Hz (2011 outburst) than its previous day. This is clearly an indication of the beginning of the declining HIMS.  

\noindent{\it (vi) Hard-Intermediate State in Declining phase:} In both the outbursts, the source enters in this state right after 
the occurrences of the maximum value of QPO frequency and remains in this state showing a prominent Type C
QPO almost on a daily basis. In this phase, the disk rate drops rapidly and the shock moves away from the black hole, 
thereby increasing the size of the Compton cloud. Due to lower cooling, the strength of 
the shock also increases. Due to reduction of viscosity at the outer disk, which causes the initiation of the declining state, 
the Keplerian rate continues to decrease and the halo rate increases (Fig. 2c, Appendix Table I for 2010 and Fig. 4c, Appendix Table II for 2011). 
The spectrum makes a transition on MJD = $55462.6$ for 2010 and on MJD = $55695.4$ for 2011 from the HIMS to the
HS showing a sudden jump in halo rate. The observed halo rate is found to be maximum due to a sudden rise in the location 
and the strength of the shock.
  
\noindent{\it (vii) Hard State in Declining phase:} This state starts with a high halo rate in both the outbursts and 
continued from MJD = $55462.6$ and from MJD = $55695.4$ for 2010 and 2011 outburst till end of the observation. 
During this period, the Keplerian rate is very low with a comparatively higher halo rate.
The shock moves away with increasing strength (see Fig. 2, Fig. 4 and Appendix Table I, Appendix Table II). 
QPO was observed till the last date ($0.74$~Hz to $79$ mHz in 2010 and $1.186$~Hz to $0.382$ Hz in 2011).

\subsection{Mass Estimation using TCAF solution}

We applied Molla et al. (2016), constant TCAF model normalization method to estimate mass of the source by analyzing 
2010 and 2011 outbursts of H~1743-322. According to TCAF solution, model normalization (N) should not vary significantly on a daily basis 
of a particular BH binary system, since it depends on physical intrinsic parameters, such as, mass, distance, disk inclination angle etc.
So, we applied this concept to estimate mass of the BHC H~1743-322 by combined spectral analysis of the source during its 
two successive outbursts 2010 and 2011 using RXTE/PCA data. We first fitted the spectra using current version of the TCAF solution 
by keeping all model parameters as free. Interestingly, obtained model normalization values comes within a narrow range for both the 
outbursts across all the spectral states. During 2010 outburst N is found within $9.79 - 27.36$, and for 2011 outburst, it is found in between $10.01 - 16.18$. 
Since, in TCAF solution, mass is an independent parameter, we obtained mass range of the source as $9.51 M_\odot - 12.42 M_\odot$ 
for 2010 outburst and $9.43 M_\odot - 12.71 M_\odot$ for 2011 outburst. We then refitted all the spectra using constant 
normalization value N=15.55 (obtained by taking average of the free model normalization values of both outbursts) and 
we obtained mass range of the source as $9.63 {M}_\odot$ to $12.34 {M}_\odot$ (for 2010 outburst), and 
$9.25 {M}_\odot$ to $12.86 {M}_\odot$ (for 2011 outburst). Now, by combining the results of 2010 \& 2011 outbursts, 
the mass of the black hole candidate source can be constrained within a range of $9.25 {M}_\odot$ to $12.86 {M}_\odot$.
To verify this estimated mass range of the source, we refit all spectra by freezing all the model parameters. Mass 
of the BH was changed in $9 M_\odot - 13 M_\odot$ range to observe variation of $\chi^2_{red}$. We also restrict ourself $\chi^2_{red} \leq 2$ 
for best fits. The parabolic variations of the $\chi^2_{red}$ (as shown in Fig. 7 \& 8) for all spectra show minima in between $\sim 
10.6 M_\odot - 11.6 M_\odot$.

\subsection{Mass Estimation using Photon Index ($\Gamma$) - QPO frequency ($\nu$) Correlation Method}

Titarchuk \& Fiorito (2004) introduced a model to correlate between spectral fitted photon index ($\Gamma$) and the observed 
QPO frequency ($\nu$). This method is used to estimate mass of a few BHCs (ST07). In this model, the central source is 
surrounded by a `Compton cloud' along with a transition layer between this Compton cloud and Keplerian disk. According to 
this model the change in QPO frequency explained as the magneto-acoustic resonance oscillation frequency of the bounded 
transition layer (Titarchuk \& Wood, 2002) and the photon index (due to change in optical depth of the hot electrons) of 
emitted spectrum is due to change in the size of Compton cloud. In the TCAF paradigm (Chakrabarti \& Titarchuk, 1995) this 
Compton cloud is the post-shock region of the accretion disk and the oscillation occurs due to near equality of Compton cooling
time scale and the infall time scale (MSC96; CMD15). The change in QPO frequency and the photon index is also demonstrated 
in TCAF solution (Chakrabarti et al. 2008; Nandi et al. 2012; Debnath et al. 2014). 
The empirical relation given in ST07 by,
$$
\Gamma(\nu) = A - D B ln[exp(\frac{\nu_{tr} - \nu} {D}) +1]
\eqno{(1)}
$$
where $A$ is the saturation point of photon index, $B$ is the slope of QPO - PL index curve which scales the mass and  $\nu_{tr}$ is the 
transition/threshold frequency above which the saturation of $\Gamma$ happens. Value of $D$ controls how fast the transition occurs. 
In order to find mass of an unknown source BH, this method requires another source of known mass BH with similar type of correlation curve
as a reference.We consider the 2005 outburst data of GRO~J1655-40 as the reference source for scaling of PL index - QPO frequency
correlations to estimate BH mass in H 1743-322, because photon indices for these two objects demonstrate the same saturation level (see Fig.9).
Note, this is necessary condition for applicability of scaling technique (see details in Sec. 7.3 of Seifina et al. 2014). This outburst data  
was also used by ST07 to calculate mass of Cyg X-1 and use the same values of A, B, D and $\nu_{tr}$ as used by ST07 to fit the data of 
GRO~J1655-40. In order to study the correlation between QPO and PL-index we extract PL index using CompTB model (Farinelli et al. 2008) and we 
found that the value of PL index obtained from CompTB model is slightly higher due to bulk motion Comptonization effect which hardens the 
spectrum a little bit than the PL index obtained from DBB - PL model, this difference in PL index would cause the difference in the value of mass of 
central object. During the fitting we freeze D = $1.0$ as in ST07. We use 2005 outburst data of GRO~J1655 - 40 and use the fitted values 
of $A_{J1655} = 2.28 \pm 0.07$, $B_{J1655} = 0.13 \pm 0.02 Hz^{-1}$ and $\nu_{tr,J1655} = 6.64 \pm 0.51 Hz$ as used in ST07 paper. 
The best fitted curve is shown by solid blue (online) curve in Fig. 9. The best fitted value for H~1743-322 is $A_{H1743} = 2.23 \pm 0.05$, 
$B_{H1743} = 0.246 \pm 0.035 Hz^{-1}$ and $\nu_{tr,H1743} = 2.95 \pm 0.58 Hz$. The best fitted curve is shown by solid red (online) line in 
Fig. 9. In this figure monotonical growth and elongated part with index saturation (constancy) on the level of $\Gamma$ = 2.23 are well seen. 
The index revealed in H1743-322 based on 2010 and 2011 data is similar to those established in a number of other black hole candidates 
and has been considered to be an observational evidence for the presence of a black hole in this source (see, ST07, ST09 and references therein). 
Applying ST07 method for BH source H~1743-322 by taking GRO~J1655-40 as a reference source of mass $6.3 \pm 0.5 M_\odot$ (Greene et al. 2001) 
we obtain mass of this BH source as, 
$$
M_{H1743} = B_{H1743}\frac{M_{J1655}} {B_{J1655}} = 11.65 \pm 0.67 M_\odot
$$
which is in the same ball park as in our result.

\section{Discussions and Concluding Remarks}

In this paper we estimated the mass of the black hole candidate H~1743-332 by using two independent methods.
First, we investigate the evolution of spectral properties of Galactic transient black hole candidate H~1743-322 
during its 2010 \& 2011 outbursts keeping the normalization as a constant parameter. We extract mass of the central object 
independently from each observation as was done in Molla et al. (2016). We use the TCAF solution 
after its inclusion as a local additive table model in HEARSRC's spectral analysis software package XSPEC. 
Depending on the extracted flow parameters and nature of QPOs, similar to Debnath et al. (2013), we also observe 
four spectral states, such as, hard, hard-intermediate, soft-intermediate and soft during both the 2010 \& 2011 
outbursts of H~1743-322. In the process of fitting, we found that the model normalization remains within a narrowest 
possible range for acceptable fits. Most importantly, we also found that the same average normalization obtained 
from the 2010 outburst fits the data of 2011 outburst equally well. In order to obtain a rough estimation 
of the variation of thermal (DBB) and non-thermal (PL) fluxes during the outbursts, we follow the procedure 
given in Debnath et al. (2013) and see that QPO is totally absent in soft states for both 
the outbursts as in 2003 outburst of H~1743-322 (McClintock et al., 2009; ST09). 
It is to be noted that low frequency QPOs are thought to be due to shock oscillation phenomenon in the TCAF paradigm. Shocks
oscillate when the cooling time scale roughly matches with the infall time scale (MSC96; CMD15) inside CENBOL.
Oscillating shock intercepts oscillating number of soft photons and hard photons are emitted accordingly, causing 
QPOs (Chakrabarti \& Manickam, 2000).

In order to keep the normalization constant or nearly constant, for a better fit, sometimes we require two Gaussian curves 
whose peaks are found to be at around $3.8$~keV and $6.5$~keV (see, dotted curves in Figs. 6b, 6d and short dashed curves in Figs. 6a, 6c). 
We interpret $6.5$~keV line as the iron-line emitted from the Keplerian disk. However, the peak around $3.8$~keV line 
is clearly not the red-shifted component of the iron line, because it is much stronger than the $6.5$~keV line. 
Moreover, it is only about one-fifth of the continuum photon flux. The only explanation of this component must 
be that it is the signature of reappearance of the inner Keplerian disk as the soft-state is approached. Indeed, it can be seen 
from the Appendix Tables I and II that the normalization of this component is rising as one goes from the harder 
state to the softer state and is going down again in the declining phase. In any case, since the pressure and therefore shear is 
higher at the inner edge, the angular momentum transport is fastest (Chakrabarti \& Molteni, 1995) and the reappearance of the 
Keplerian disk in softer states must commence from the inner disk. So our work possible points to this aspect very clearly. 

Spectral analysis of 2010 outburst of this source with TCAF model has been done by Mondal et al. (2014) assuming 
a fixed mass of the black hole ($11.4 M_\odot$) and keeping the model normalization free. 
On the other hand, to determine the mass independently, we fitted all the spectra using 
all TCAF model parameters as free and for both outbursts, we observe a narrow variation of model normalization 
($9.79 - 27.36$ for 2010 outburst and $10.01 - 16.18$ for 2011 outburst). 
Mass of the source is obtained in the range: $9.51 M_\odot - 12.42 M_\odot$ (for 2010 outburst) 
and $9.43 M_\odot - 12.71 M_\odot$ (for 2011 outburst). We then refitted all the spectra using constant
average normalization value of $N=15.55$ and obtained the mass from $9.63 {M}_\odot$ to $12.34 {M}_\odot$ (for 2010 outburst), and
$9.25 {M}_\odot$ to $12.86 {M}_\odot$ (for 2011 outburst). We then computed variation of $\chi^2_{red}$ with mass in this general 
range of $9-13 M_\odot$ and found that $\chi^2_{red}$ is minimum in between $\sim 10.6 M_\odot - 11.6 M_\odot$. 
To compare, we also estimate the mass of the source using another independent method, i.e., ST07 photon index ($\Gamma$) 
vs. QPO frequency correlation scaling) by taking data of the both 2010 and 2011 outbursts of H~1743-322. 
Our estimated mass from this method is $11.65 \pm 0.67 M_\odot$, which was obtained by using GRO~J1655-40 
as a reference source. Combining both the TCAF and the ST07 methods, we believe that the mass of H~1743-322 should be 
in the range of $11.21^{+1.65}_{-1.96} M_\odot$. We find that the TCAF result gives the mean mass to be 
slightly lower as compared to the mean mass from ST09.

TCAF represents a comprehensive solution of equations governing the flow. It attempts to explain the spectral and 
timing properties simultaneously and extract physical parameters such as the accretion rates, size 
of the Compton cloud and even the Mass of the central object. Each individual observation gives a new and independent 
mass estimate. This is in contrast with the method of ST07 and ST09 where a refereence source is also required. So far, 
the effects of the magnetic field and spin have not been included in TCAF. However, since shocks are located far away 
in harder states, effects of spin would not be significant on the mass obtained from harder states. Nevertheless, 
we are incorporating these effects and the results would be published elsewhere.


\section*{Acknowledgments}

A.A.M. and S.M. acknowledge supports from MoES sponsored senior and post-doctoral research fellowships respectively.
D.D. acknowledges support from project fund of DST sponsored Fast-track Young Scientist (SR/FTP/PS-188/2012).
We thank the anonymous referee of this paper for very detailed and useful comments, which helped in improving the quality of the paper.

{}

\clearpage

\begin{figure}
\vskip -0.5cm
\centerline{
\includegraphics[scale=0.6,angle=270,width=8.0truecm]{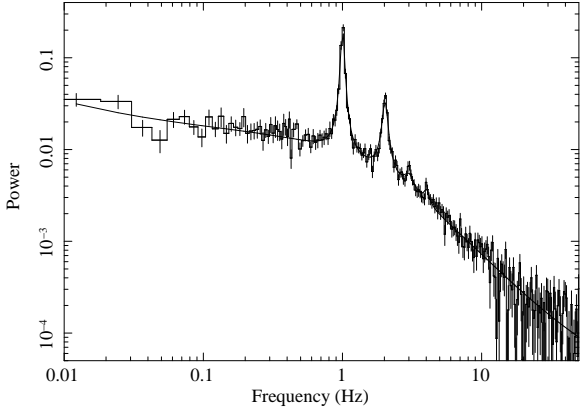}}
\caption{Power density spectrum of $0.01$~sec time binned PCU2 light curve in the energy of $2-15$~keV with an
exposure time of 3379s is shown for the observation Id 95360-14-01-00 (MJD=55418.43) of 2010 outburst. 
It has been fitted with combination of five Lorentzians (one broad for initial break part and others for 
four QPO peaks) and one power-law (for late slope) models. The frequency of the primary dominating QPO is found to be $1.008 \pm 0.003 $~Hz.}
\label{fig1}
\end{figure}

\begin{figure}
\vskip -0.5cm
\centerline{
\includegraphics[scale=0.6,angle=0,width=8.0truecm]{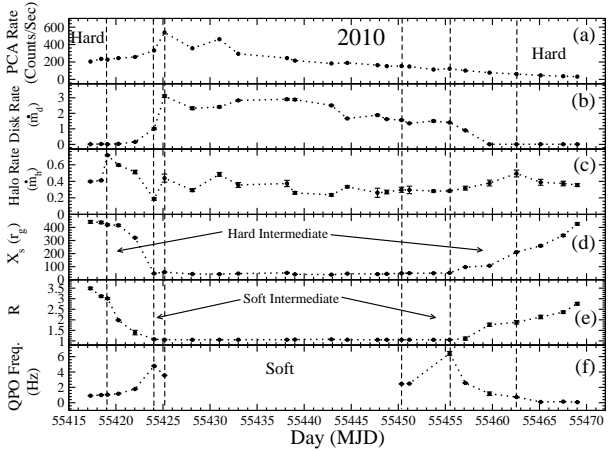}}
\caption{Variation of (a) $2.5-25$~keV PCA count rates (cnts/sec), (b) TCAF fitted disk (accretion) rate $\dot{m_d}$
in $\dot{M}$$_{Edd}$ unit, (c) sub-Keplerian halo $\dot{m_h}$ rate in $\dot{M}$$_{Edd}$ unit, and (d) TCAF fitted 
shock location ($X_s$) in $2GM/c^2$ unit, (e) strength of the shock $(R)$, in the $2.5-25$~keV energy band keeping an average normalization 
constant of $N=15.55$. Along the X-axis is the observation day (MJD) for the 2010 outburst. In (f), the
observed primary dominating QPO frequencies (in Hz) with day (MJD) are shown. The vertical dashed lines indicate 
transitions between different spectral states.}
\label{fig2}
\end{figure}

\begin{figure}
\vskip -0.5cm
\centerline{
\includegraphics[scale=0.6,angle=0,width=8.0truecm]{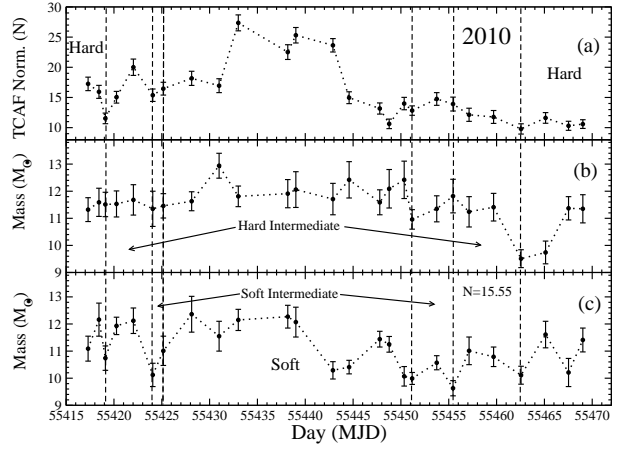}}
\caption{Variations of TCAF model fitted parameters: (a) normalization, and (b) mass of the black hole 
($M_{BH}$ in $M_\odot$) over the entire period of the 2010 outburst of H~1743-322 are shown when all model 
parameters are kept as free. In (c), TCAF fitted $M_{BH}$ values are shown when model normalization was kept frozen at 
$15.55$, which we get by taking the weighted average of the normalization values of panel (a).}
\label{fig3}
\end{figure}

\begin{figure}
\vskip -0.5cm
\centerline{
\includegraphics[scale=0.6,angle=0,width=8.0truecm]{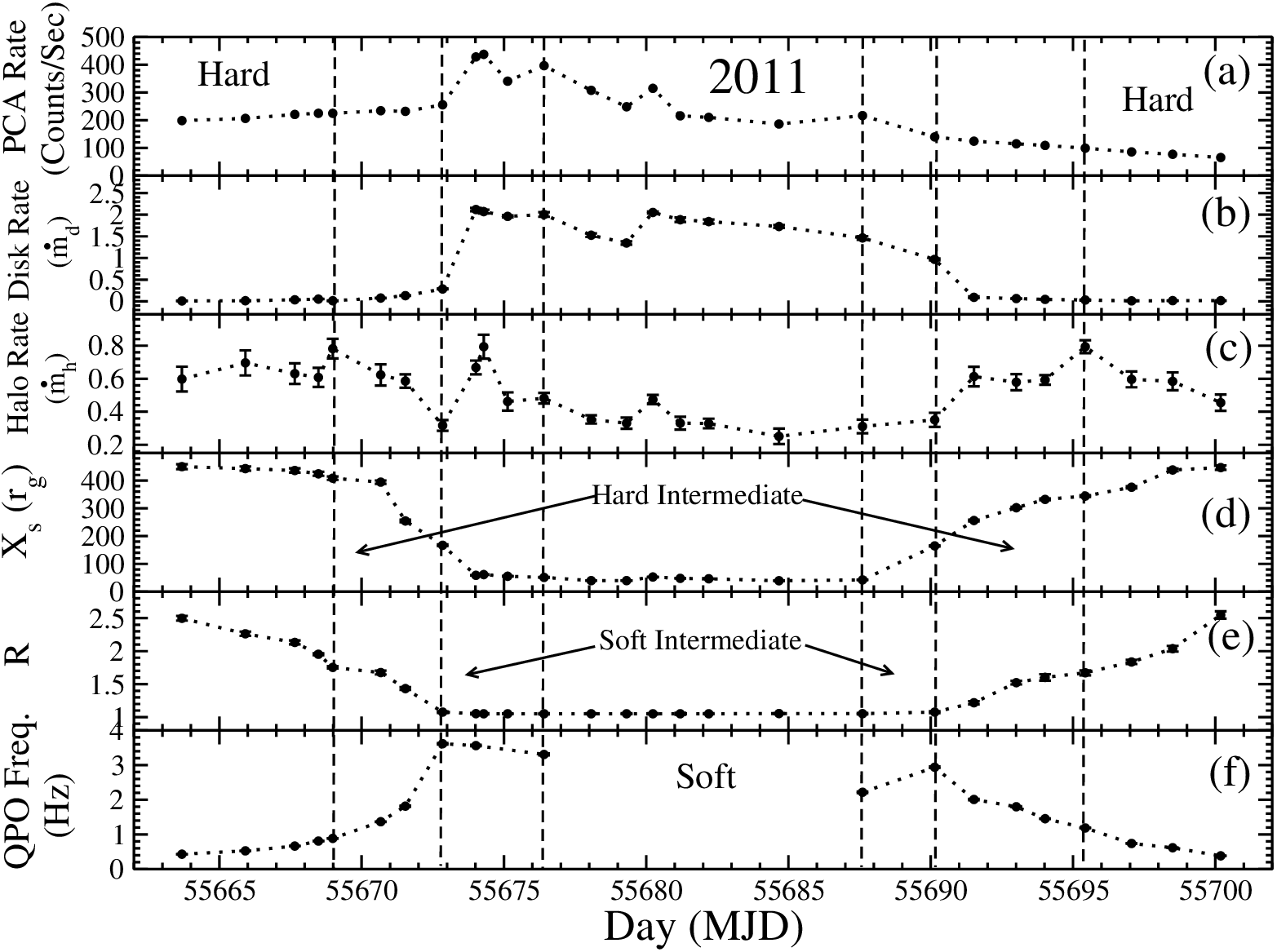}}
\caption{Same as Fig. 2, except for 2011 outburst H~1743-322}
\label{fig4}
\end{figure}

\begin{figure}
\vskip -0.5cm
\centerline{
\includegraphics[scale=0.6,angle=0,width=8.0truecm]{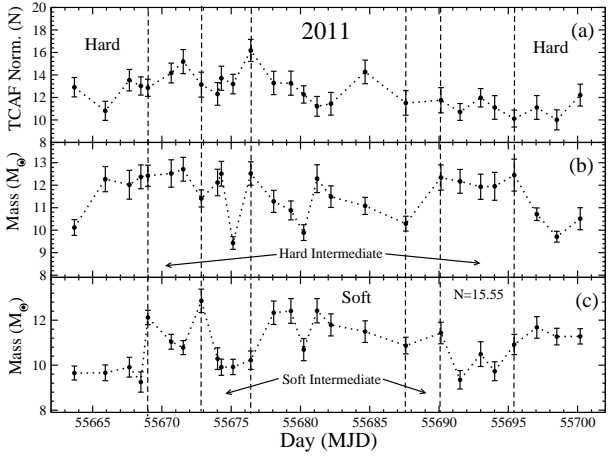}}
\caption{Same as Fig. 3, except for 2011 outburst of H~1743-322.}
\label{fig5}
\end{figure}

\begin{figure}
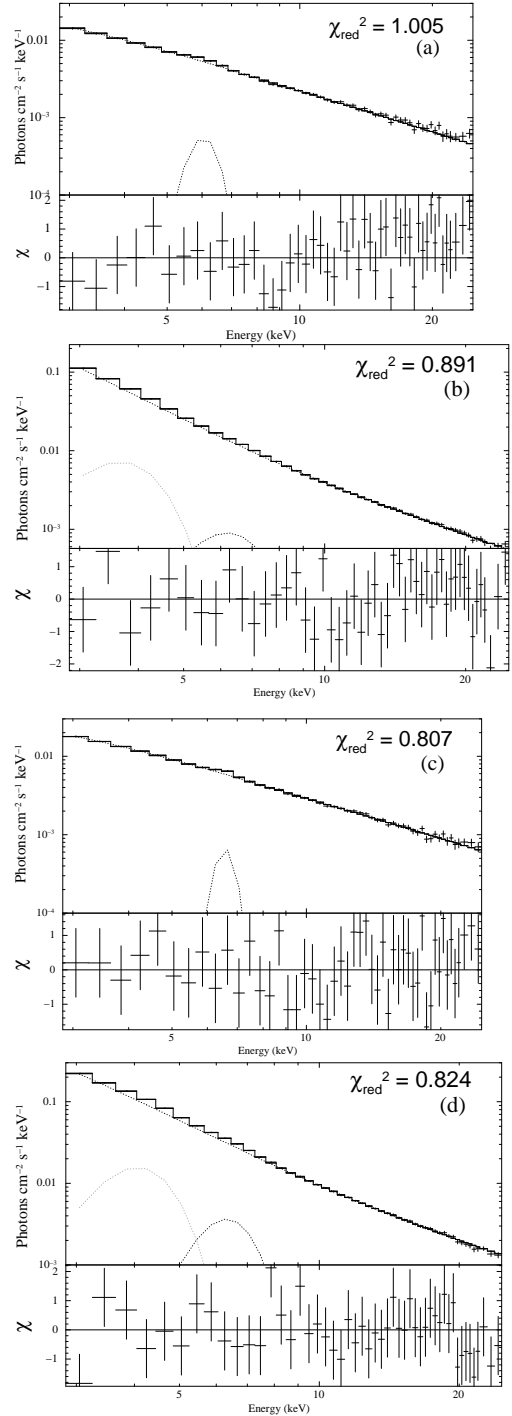

\includegraphics[width=0.62\columnwidth,angle=270]{fig6a.ps} 
\includegraphics[width=0.62\columnwidth,angle=270]{fig6b.ps}
\includegraphics[width=0.62\columnwidth,angle=270]{fig6c.ps}
\includegraphics[width=0.62\columnwidth,angle=270]{fig6d.ps}

\caption{TCAF model fitted unfolded spectrum for (a) hard state (Obs Id: 95360-14-25-01), (b) soft state (Obs Id: 95360-14-15-00) from 
2010 outburst of H~1743-322 are shown. Same for (c) hard state (Obs Id: 96425-01-06-02), (d) soft state (Obs Id: 96425-01-03-00) from
2011 outburst of H~1743-322 are shown. For a constant normalization, better fits, especially in softer states require double Gaussians
having peaks at around 3.8 keV and $6.5$ keV respectively with the normalization of the first component increasing with photon index. 
See text for details.}

\label{fig6}
\end{figure}

\begin{figure}
\vskip -0.5cm
\centerline{
\includegraphics[scale=0.6,angle=0,width=8.0truecm]{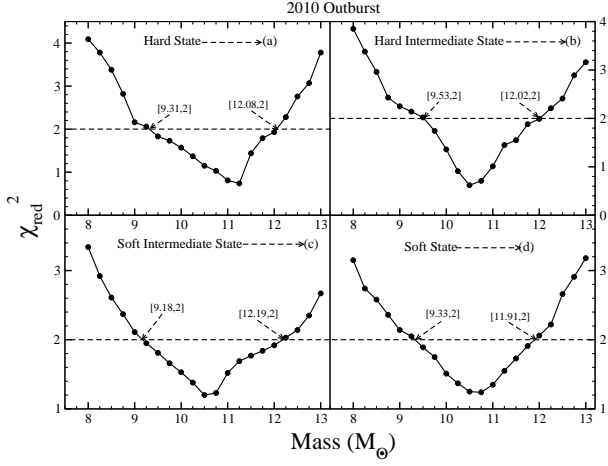}}
\caption{(a)-(d) variation of $\chi^2_{red}$ with mass of the black hole in solar 
mass($M_\odot$) unit for four different observations selected from four spectral 
states HS, HIMS, SIMS and SS for 2010 outburst. Here the square brackets denotes the value of 
mass for acceptable reduced $\chi^2$ limit ($\le 2.0$) .}
\label{fig7}
\end{figure}

\begin{figure}
\vskip -0.5cm
\centerline{
\includegraphics[scale=0.6,angle=0,width=8.0truecm]{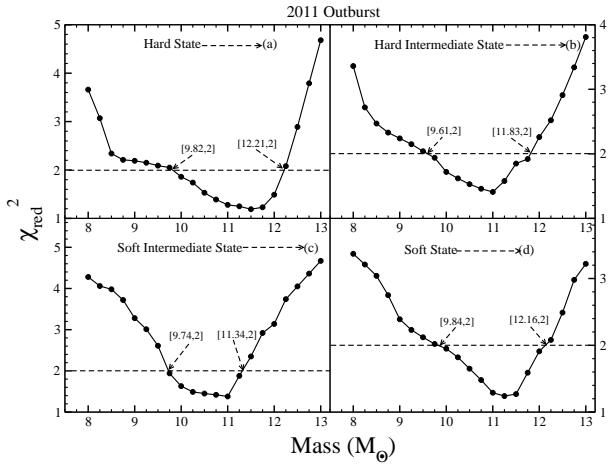}}
\caption{Same as Fig. 7, except for 2011 outburst of H 1743-322}
\label{fig8}
\end{figure}

\begin{figure}
\vskip -0.5cm
\centerline{
\includegraphics[scale=0.6,angle=0,width=8.0truecm]{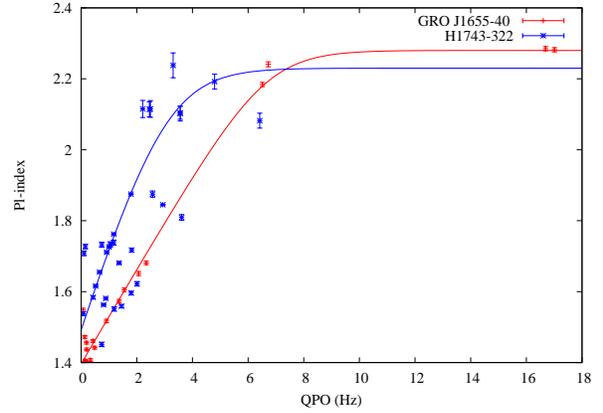}}
\caption{Photon index - QPO frequency correlations data for H~1743-322 for 2010 \& 2011 outbursts (online blue points) 
and GRO~J1655-40 during the 2005 outburst data (online red points). The data are fitted with a model function to apply 
a scaling method. We obtain the BH mass H~1743-322 in the range of $11.65 \pm 0.67 M_\odot$ using the scaling method 
and the well-known BH mass of GRO~J1655-40, which is in the range of $6.3 \pm 0.5 M_\odot$. The result is obtained by using
model composition wabs$\times$(CompTB+Gaussian).}
\label{fig9}
\end{figure}

\clearpage

\begin{table}
  \vskip -2.0cm
\addtolength{\tabcolsep}{-4.0pt}
\scriptsize
\centering
\centering{\large \bf Appendix Table I}
\vskip 0.2cm
\centerline {\normalsize TCAF Model Fitted parameters during 2010 outburst}
\vskip 0.2cm
	\begin{tabular}{lccccccccccccccc}
	\hline
  Obs. & Id& Day & $\dot{m_d}$  & $\dot{m_h}$& $X_S$& $R$ &$ga1$ &$sigma1$&$norm1$&$ga2$&$sigma2$&$norm2$&$QPO$&$\chi^2_{red}$(dof) \\ 
       &   &(MJD)&($\dot{M}$$_{Edd}$)& ($\dot{M}_{Edd}$) & $r_g$  &  &$(keV)$& $(keV)$&$*10^{-3}$ &$(keV)$&$(keV)$&$*10^{-3}$&$(Hz)$ \\ 
  (1)  & (2)& (3) &  (4)              &(5) &  (6)   & (7)& (8)   &  (9)   & (10)  &  (11)& (12)  & (13)  & (14)  &  (15)              \\ 
\hline
 
 1& X-01-00 & 55417.3 & $0.013^{\pm 0.005}$ & $0.398^{\pm 0.007}$ & $442.1^{\pm 10.6}$ &$3.491^{\pm 0.056}$&$6.31^{\pm 0.26}$&$0.65^{\pm 0.15}$&$2.9^{\pm 0.8}$&$3.83^{\pm 0.21}$&$0.79^{\pm 0.25}$&$4.58^{\pm 1.11}$& $0.919^{\pm 0.004}$ & 1.09(39) \\
 2& Y-01-00 & 55418.4 & $0.025^{\pm 0.009}$ & $0.411^{\pm 0.005}$ & $437.7^{\pm 9.3}$ & $3.117^{\pm 0.045}$&$6.39^{\pm 0.31}$&$0.58^{\pm 0.21}$&$2.5^{\pm 0.6}$&$3.89^{\pm 0.25}$&$0.81^{\pm 0.18}$&$4.21^{\pm 1.32}$& $1.008^{\pm 0.003}$ & 0.76(39)\\
 3& Y-02-01 & 55419.1 & $0.011^{\pm 0.005}$ & $0.715^{\pm 0.007}$ & $419.7^{\pm 6.3}$ & $3.012^{\pm 0.039}$&$6.22^{\pm 0.38}$&$0.81^{\pm 0.24}$&$4.8^{\pm 1.2}$&$3.59^{\pm 0.17}$&$0.76^{\pm 0.23}$&$7.91^{\pm 1.55}$&$1.045^{\pm 0.008}$ &  1.51(39)\\
 4& Y-02-00 & 55420.3 & $0.039^{\pm 0.011}$ & $0.597^{\pm 0.011}$ & $415.6^{\pm 7.9}$ & $1.981^{\pm 0.015}$&$6.22^{\pm 0.35}$&$0.75^{\pm 0.14}$&$3.7^{\pm 0.8}$&$3.66^{\pm 0.38}$&$0.82^{\pm 0.13}$&$6.11^{\pm 1.75}$&$1.174^{\pm 0.002}$& 1.26(39)\\
 5& Y-02-03 & 55422.0 & $0.158^{\pm 0.019}$ & $0.612^{\pm 0.021}$ & $320.9^{\pm 3.1}$ & $1.394^{\pm 0.107}$&$6.54^{\pm 0.45}$&$0.31^{\pm 0.09}$&$2.1^{\pm 0.9}$&$4.42^{\pm 0.42}$&$0.35^{\pm 0.11}$&$2.08^{\pm 1.12}$& $1.789^{\pm 0.017}$ & 0.68(39)\\
 6& Y-03-01 & 55424.1 & $1.008^{\pm 0.051}$ & $0.186^{\pm 0.015}$ & $ 47.6^{\pm 1.2}$ & $1.075^{\pm 0.015}$&$6.13^{\pm 0.32}$&$0.75^{\pm 0.16}$&$5.6^{\pm 1.6}$&$3.35^{\pm 0.35}$&$0.65^{\pm 0.21}$&$29.1^{\pm 5.4}$&$4.796^{\pm 0.022}$ & 1.19 (39)\\
 7& Y-04-00 & 55425.2 & $3.105^{\pm 0.084}$ & $0.439^{\pm 0.049}$ & $ 58.3^{\pm 2.2}$  & $1.058^{\pm 0.013}$&$6.08^{\pm 0.18}$&$0.92^{\pm 0.13}$&$17.1^{\pm 2.3}$&$3.91^{\pm 0.49}$&$0.81^{\pm 0.15}$&$64.7^{\pm 8.6}$& $3.558^{\pm 0.024}$ & 1.39(39)\\
 8& Y-05-00 & 55428.1 & $2.324^{\pm 0.085}$ & $0.273^{\pm 0.017}$ & $ 43.8^{\pm 1.5}$  & $1.053^{\pm 0.005}$&$6.16^{\pm 0.29}$&$0.95^{\pm 0.12}$&$ 9.2^{\pm 1.6}$&$3.85^{\pm 0.72}$&$0.81^{\pm 0.28}$&$57.9^{\pm 9.8}$& $*****$           & 1.33(39)\\
 9& Y-07-00 & 55431.1 & $2.412^{\pm 0.058}$ & $0.484^{\pm 0.023}$ & $ 42.8^{\pm 1.2}$  & $1.053^{\pm 0.004}$&$6.37^{\pm 0.39}$&$0.71^{\pm 0.11}$&$6.7^{\pm 1.9}$&$3.98^{\pm 0.47}$&$0.77^{\pm 0.25}$&$42.4^{\pm 7.8}$  & $*****$           & 1.39(39)\\
10& Y-06-01 & 55433.1 & $2.831^{\pm 0.033}$ & $0.356^{\pm 0.029}$ & $ 47.2^{\pm 0.9}$  & $1.054^{\pm 0.005}$&$6.29^{\pm 0.45}$&$0.55^{\pm 0.15}$&$5.4^{\pm 1.4}$&$3.31^{\pm 0.32}$&$0.88^{\pm 0.21}$&$84.2^{\pm 5.1}$& $*****$           &  0.67(39)\\
11& Y-10-00 & 55438.2 & $2.902^{\pm 0.058}$ & $0.374^{\pm 0.038}$ & $ 52.5^{\pm 1.5}$  & $1.054^{\pm 0.004}$&$6.21^{\pm 0.29}$&$0.81^{\pm 0.24}$&$3.3^{\pm 0.6}$&$3.62^{\pm 0.46}$&$0.74^{\pm 0.32}$&$40.6^{\pm 8.4}$& $*****$           &  0.87(39)\\
12& Y-11-00 & 55439.1 & $2.881^{\pm 0.061}$ & $0.261^{\pm 0.016}$ & $ 41.5^{\pm 0.7}$  & $1.062^{\pm 0.004}$&$6.11^{\pm 0.34}$&$0.71^{\pm 0.19}$&$4.0^{\pm 1.1}$&$3.15^{\pm 0.23}$&$0.95^{\pm 0.08}$&$101^{\pm 21}$& $*****$            &  0.92(39)\\
13& Y-13-00 & 55442.9 & $2.508^{\pm 0.028}$ & $0.236^{\pm 0.015}$ & $ 39.1^{\pm 0.7}$  & $1.071^{\pm 0.005}$&$5.91^{\pm 0.44}$&$0.98^{\pm 0.07}$&$4.4^{\pm 1.7}$&$3.32^{\pm 0.36}$&$0.88^{\pm 0.14}$&$63.7^{\pm 9.8}$  & $*****$          & 1.25(39)\\
14& Y-15-00 & 55444.6 & $1.668^{\pm 0.019}$ & $0.334^{\pm 0.013}$ & $ 46.9^{\pm 0.6}$  & $1.055^{\pm 0.003}$&$6.21^{\pm 0.41}$&$0.91^{\pm 0.09}$&$2.2^{\pm 0.8}$&$3.62^{\pm 0.55}$&$0.69^{\pm 0.24}$&$14.5^{\pm 3.5}$& $*****$            & 0.89(39)\\
15& Y-18-00 & 55447.8 & $1.885^{\pm 0.048}$ & $0.262^{\pm 0.055}$ & $ 43.2^{\pm 0.6}$  & $1.053^{\pm 0.002}$&$6.22^{\pm 0.45}$&$0.75^{\pm 0.15}$&$2.1^{\pm 0.7}$&$3.38^{\pm 0.18}$&$0.78^{\pm 0.32}$&$19.7^{\pm 4.8}$& $*****$            & 1.14(39)\\
16& Y-19-00 & 55448.8 & $1.623^{\pm 0.022}$ & $0.271^{\pm 0.048}$ & $ 44.9^{\pm 0.8}$  & $1.052^{\pm 0.004}$&$6.21^{\pm 0.43}$&$0.85^{\pm 0.13}$&$1.6^{\pm 0.5}$&$3.21^{\pm0.23}$&$0.42^{\pm 0.21}$&$27.1^{\pm 5.4}$& $*****$             & 0.95(39)\\
17& Y-20-00 & 55450.3 & $1.564^{\pm 0.041}$ & $0.298^{\pm 0.031}$ & $ 49.1^{\pm 0.9}$  & $1.053^{\pm 0.003}$&$6.62^{\pm 0.51}$&$0.75^{\pm 0.31}$&$2.8^{\pm 0.9}$&$4.19^{\pm 0.39}$&$0.32^{\pm 0.16}$&$1.33^{\pm 0.45}$& $2.454^{\pm 0.022}$ &  0.61(39)\\
18& Y-20-01 & 55451.2 & $1.355^{\pm 0.025}$ & $0.296^{\pm 0.045}$ & $ 50.2^{\pm 1.1}$  & $1.054^{\pm 0.003}$&$6.39^{\pm 0.48}$&$0.85^{\pm 0.35}$&$2.9^{\pm 0.9}$&$3.73^{\pm 0.33}$&$0.63^{\pm 0.21}$&$4.12^{\pm 1.11}$& $2.489^{\pm 0.018}$ &  0.59(39)\\
19& Y-21-01 & 55453.7 & $1.503^{\pm 0.055}$ & $0.282^{\pm 0.012}$ & $ 49.9^{\pm 1.1}$  & $1.052^{\pm 0.004}$&$6.51^{\pm 0.45}$&$0.31^{\pm 0.11}$&$0.3^{\pm 0.1}$&$3.18^{\pm 0.22}$&$0.59^{\pm 0.34}$&$11.8^{\pm 2.1}$& $*****$           &  0.81(39)\\
20& Y-22-01 & 55455.4 & $1.408^{\pm 0.019}$ & $0.284^{\pm 0.011}$ & $ 52.5^{\pm 1.9}$  & $1.053^{\pm 0.003}$&$6.71^{\pm 0.52}$&$0.71^{\pm 0.18}$&$2.1^{\pm 0.8}$&$4.44^{\pm 0.56}$&$0.31^{\pm 0.14}$&$0.27^{\pm 0.07}$& $6.417^{\pm 0.252}$ &  0.86(39)\\
21& Y-23-01 & 55457.1 & $0.041^{\pm 0.009}$ & $0.311^{\pm 0.025}$ & $ 96.1^{\pm 2.2}$  & $1.136^{\pm 0.085}$&$6.56^{\pm 0.62}$&$0.33^{\pm 0.16}$&$0.1^{\pm 0.1}$&$----$&$----$&$----$& $2.569^{\pm 0.044}$ & 1.09(42)\\
22& Y-24-01 & 55459.7 & $0.011^{\pm 0.005}$ & $0.379^{\pm 0.035}$ & $107.3^{\pm 3.1}$  & $1.764^{\pm 0.077}$&$6.37^{\pm 0.12}$&$0.32^{\pm 0.08}$&$0.2^{\pm 0.1}$&$----$&$----$&$----$& $1.172^{\pm 0.226}$ & 1.31(42)\\
23& Y-25-01 & 55462.6 & $0.010^{\pm 0.005}$ & $0.495^{\pm 0.041}$ & $211.3^{\pm 4.9}$  & $1.882^{\pm 0.085}$&$6.04^{\pm 0.14}$&$0.42^{\pm 0.22}$&$0.1^{\pm 0.1}$&$----$&$----$&$----$& $0.741^{\pm 0.046}$ & 1.01(42)\\
24& Y-28-00 & 55465.1 & $0.011^{\pm 0.004}$ & $0.388^{\pm 0.036}$ & $259.3^{\pm 5.9}$  & $2.134^{\pm 0.082}$&$6.63^{\pm 0.55}$&$0.33^{\pm 0.18}$&$0.3^{\pm 0.1}$&$----$&$----$&$----$& $0.102^{\pm 0.003}$ & 1.24(42)\\
25& Y-26-02 & 55467.5 & $0.010^{\pm 0.005}$ & $0.374^{\pm 0.028}$ & $338.3^{\pm 7.7}$  & $2.365^{\pm 0.055}$&$6.42^{\pm 0.41}$&$0.35^{\pm 0.09}$&$0.1^{\pm 0.1}$&$----$&$----$&$----$& $0.149^{\pm 0.005}$ & 1.02(42)\\
26& Y-28-01 & 55469.0 & $0.011^{\pm 0.004}$ & $0.355^{\pm 0.017}$ & $426.6^{\pm 9.2}$  & $2.762^{\pm 0.062}$&$6.41^{\pm 0.29}$&$0.74^{\pm 0.17}$&$0.1^{\pm 0.1}$&$----$&$----$&$----$& $0.079^{\pm 0.002}$ & 1.11(42)\\
\hline
\end{tabular}
\vskip 0.4cm
\noindent{Here X=95368-01, Y=95360-14 are the initial part of observation IDs. In Columns 4, 5, 6 \& 7 we show TCAF} 
\noindent{model fitted disk rate ($\dot{m}_d$), halo rate ($\dot{m}_h$), shock location ($X_s$) and shock compression}
\noindent{ratio ($R$) along with their error bars during the 2010 outburst. In Columns 8,9,10,11,12 \& 13 we show Gaussian}
\noindent{line energy, sigma and normalization for two Gaussian used to fit the spectra of the BH} 
\noindent{In Columns 14 \& 15 we present observed QPO values and value of $\chi^2_{red}$ with no. of degrees} 
\noindent{of freedom mentioned in the brackets. The ``err" command is used to find $90\%$ confidence $\pm$ error} 
\noindent{values for the model fitted parameters. The result is obtained by using model composition wabs*(TCAF+Gaussian+Gaussian).}
 
\end{table}

\clearpage

\begin{table}
\vskip -2.0cm
\addtolength{\tabcolsep}{-4.0pt}
\scriptsize
\centering
\centering{\large \bf Appendix Table II}
\vskip 0.2cm
\centerline {\normalsize TCAF Model Fitted parameters during 2011 outburst}
\vskip 0.2cm
	\begin{tabular}{lccccccccccccccc} 
	\hline
  Obs. & Id  & Day & $\dot{m_d}$ & $\dot{m_h}$& $X_S$& $R$& $ga1$ & $sigma1$& $norm1$& $ga2$& $sigma2$& $norm2$& $QPO$ &  $\chi^2_{red}$(dof)     \\ 
       &  & (MJD)& ($\dot{M}$$_{Edd}$)& ($\dot{M}_{Edd}$)& $r_g$  &    & $(keV)$&$(keV)$& $*10^{-3}$ &$(keV)$ &$(keV)$& $*10^{-3}$&(Hz) \\ 
 (1)   & (2) & (3) &  (4)   &  (5) &  (6) & (7) & (8) &  (9) & (10) &  (11) &  (12) & (13) & (14) & (15) \\ 
\hline
 
 1 & Z-01-00 & 55663.7 & $0.010^{\pm 0.005}$ & $0.598^{\pm 0.075}$ & $449.9^{\pm 8.8}$ & $2.498^{\pm 0.035}$&$6.41^{\pm 0.36}$&$0.47^{\pm 0.22}$&$2.1^{\pm 0.97}$&$4.02^{\pm 0.29}$&$0.77^{\pm 0.34}$&$3.9^{\pm 1.3}$ &$0.428^{\pm 0.001}$ & 0.91(39) \\
 2 & Z-01-01 & 55665.9 & $0.013^{\pm 0.007}$ & $0.696^{\pm 0.076}$ & $443.1^{\pm 7.7}$ & $2.263^{\pm 0.032}$&$6.18^{\pm 0.29}$&$0.61^{\pm 0.27}$&$2.7^{\pm 1.1}$&$3.54^{\pm 0.35}$&$0.98^{\pm 0.08}$&$7.8^{\pm 2.2}$& $0.524^{\pm 0.002}$ & 1.49(39) \\
 3 & Z-02-00 & 55667.7 & $0.035^{\pm 0.008}$ & $0.631^{\pm 0.062}$ & $436.1^{\pm 7.6}$ & $2.132^{\pm 0.029}$&$6.21^{\pm 0.22}$&$0.83^{\pm 0.11}$&$3.6^{\pm 1.2}$&$3.58^{\pm 0.32}$&$0.68^{\pm 0.21}$&$5.6^{\pm 2.4}$& $0.664^{\pm 0.002}$ & 1.25(39) \\
 4 & Z-02-04 & 55668.5 & $0.053^{\pm 0.003}$ & $0.608^{\pm 0.058}$ & $424.2^{\pm 7.1}$ & $1.953^{\pm 0.015}$&$6.31^{\pm 0.42}$&$0.61^{\pm 0.26}$&$2.1^{\pm 0.9}$&$----$&$----$&$----$& $0.807^{\pm 0.003}$ & 1.39(42) \\
 5 & Z-02-01 & 55667.0 & $0.013^{\pm 0.008}$ & $0.782^{\pm 0.059}$ & $408.5^{\pm 5.9}$ & $1.754^{\pm 0.017}$&$6.26^{\pm 0.45}$&$0.58^{\pm 0.17}$&$2.1^{\pm 0.8}$&$----$&$----$&$----$& $0.885^{\pm 0.004}$ & 1.37(42) \\
 6 & Z-02-02 & 55670.7 & $0.075^{\pm 0.007}$ & $0.623^{\pm 0.064}$ & $394.1^{\pm 6.1}$ & $1.675^{\pm 0.028}$&$6.21^{\pm 0.39}$&$0.68^{\pm 0.31}$&$3.5^{\pm 1.2}$&$3.51^{\pm 0.19}$&$0.69^{\pm 0.11}$&$5.9^{\pm 2.3}$& $1.364^{\pm 0.007}$ & 1.38(39) \\
 7 & Z-02-05 & 55671.5 & $0.131^{\pm 0.004}$ & $0.586^{\pm 0.041}$ & $254.5^{\pm 4.2}$ & $1.433^{\pm 0.021}$&$6.32^{\pm 0.42}$&$0.74^{\pm 0.22}$&$3.4^{\pm 0.8}$&$3.55^{\pm 0.32}$&$0.48^{\pm 0.14}$&$4.7^{\pm 1.6}$& $1.816^{\pm 0.015}$ & 1.40(39) \\
 8 & Z-02-03 & 55672.9 & $0.286^{\pm 0.011}$ & $0.317^{\pm 0.033}$ & $167.1^{\pm 2.9}$ & $1.076^{\pm 0.011}$&$6.23^{\pm 0.29}$&$0.72^{\pm 0.24}$&$2.1^{\pm 0.6}$&$3.33^{\pm 0.32}$&$0.45^{\pm 0.09}$&$10.8^{\pm 3.3}$& $3.614^{\pm 0.021}$ & 1.35(39) \\
 9 & Z-03-00 & 55674.0 & $2.118^{\pm 0.038}$ & $0.668^{\pm 0.042}$ & $ 58.8^{\pm 0.8}$ & $1.053^{\pm 0.009}$&$6.32^{\pm 0.46}$&$0.81^{\pm 0.12}$&$7.4^{\pm 2.9}$&$4.03^{\pm 0.41}$&$0.68^{\pm 0.21}$&$31.1^{\pm 0.1}$& $3.562^{\pm 0.019}$ & 0.82(39) \\ 
10 & Z-03-05 & 55674.3 & $2.071^{\pm 0.042}$ & $0.794^{\pm 0.072}$ & $ 61.4^{\pm 0.8}$ & $1.051^{\pm 0.007}$&$6.64^{\pm 0.52}$&$0.75^{\pm 0.18}$&$4.6^{\pm 2.1}$&$3.89^{\pm 0.49}$&$0.69^{\pm 0.21}$&$30.4^{\pm 0.1}$& $*****$           & 1.09(39) \\     
11 & Z-03-01 & 55675.1 & $1.961^{\pm 0.021}$ & $0.462^{\pm 0.055}$ & $ 55.6^{\pm 0.7}$ & $1.052^{\pm 0.004}$&$6.11^{\pm 0.53}$&$0.77^{\pm 0.28}$&$5.1^{\pm 1.9}$&$3.71^{\pm 0.41}$&$0.75^{\pm 0.29}$&$20.1^{\pm 0.1}$& $*****$           & 0.92(39) \\
12 & Z-03-02 & 55676.4 & $2.004^{\pm 0.054}$ & $0.482^{\pm 0.032}$ & $ 51.4^{\pm 1.1}$ & $1.051^{\pm 0.008}$&$6.41^{\pm 0.55}$&$0.76^{\pm 0.11}$&$6.5^{\pm 2.2}$&$3.85^{\pm 0.32}$&$0.78^{\pm 0.17}$&$40.2^{\pm 0.2}$& $3.306^{\pm 0.028}$ & 1.49(39) \\
13 & Z-03-03 & 55678.1 & $1.525^{\pm 0.041}$ & $0.353^{\pm 0.025}$ & $ 40.1^{\pm 0.6}$ & $1.052^{\pm 0.006}$&$6.24^{\pm 0.38}$&$0.55^{\pm 0.08}$&$5.6^{\pm 1.8}$&$3.55^{\pm 0.29}$&$0.45^{\pm 0.05}$&$50.1^{\pm 0.1}$& $*****$           & 1.09(39) \\       
14 & Z-03-04 & 55679.3 & $1.345^{\pm 0.039}$ & $0.331^{\pm 0.034}$ & $ 39.9^{\pm 0.7}$ & $1.052^{\pm 0.006}$&$6.02^{\pm 0.56}$&$0.72^{\pm 0.29}$&$2.8^{\pm 0.8}$&$3.35^{\pm 0.41}$&$0.65^{\pm 0.15}$&$39.8^{\pm 0.1}$& $*****$           & 1.54(39) \\       
15 & Z-04-02 & 55680.2 & $2.051^{\pm 0.021}$ & $0.474^{\pm 0.028}$ & $ 52.9^{\pm 0.8}$ & $1.051^{\pm 0.009}$&$6.03^{\pm 0.45}$&$0.78^{\pm 0.33}$&$4.9^{\pm 2.1}$&$3.79^{\pm 0.39}$&$0.67^{\pm 0.29}$&$30.2^{\pm 0.2}$& $*****$           & 1.52(39) \\       
16 & Z-04-00 & 55681.2 & $1.882^{\pm 0.052}$ & $0.331^{\pm 0.039}$ & $ 48.1^{\pm 0.7}$ & $1.051^{\pm 0.015}$&$6.05^{\pm 0.51}$&$0.76^{\pm 0.28}$&$2.7^{\pm 0.7}$&$3.43^{\pm 0.42}$&$0.54^{\pm 0.19}$&$29.9^{\pm 0.1}$& $*****$           & 0.97(39) \\     
17 & Z-04-03 & 55682.2 & $1.841^{\pm 0.048}$ & $0.329^{\pm 0.029}$ & $ 46.4^{\pm 0.5}$ & $1.052^{\pm 0.005}$&$5.05^{\pm 1.25}$&$0.58^{\pm 0.31}$&$2.7^{\pm 0.8}$&$3.41^{\pm 0.45}$&$0.62^{\pm 0.24}$&$31.1^{\pm 0.1}$& $*****$           & 0.98(39) \\
18 & Z-04-01 & 55684.7 & $1.728^{\pm 0.032}$ & $0.251^{\pm 0.047}$ & $ 39.5^{\pm 0.7}$ & $1.054^{\pm 0.006}$&$6.23^{\pm 0.45}$&$0.69^{\pm 0.22}$&$2.3^{\pm 0.5}$&$3.63^{\pm 0.31}$&$0.68^{\pm 0.23}$&$32.5^{\pm 0.1}$& $*****$           & 1.05(39) \\
19 & Z-05-00 & 55687.6 & $1.464^{\pm 0.041}$ & $0.311^{\pm 0.041}$ & $ 42.6^{\pm 0.8}$ & $1.053^{\pm 0.009}$&$6.03^{\pm 0.62}$&$0.95^{\pm 0.02}$&$3.1^{\pm 1.3}$&$3.85^{\pm 0.39}$&$0.45^{\pm 0.13}$&$6.1^{\pm 2.4}$& $2.215^{\pm 0.021}$ & 1.01(39) \\
20 & Z-05-01 & 55690.1 & $0.969^{\pm 0.022}$ & $0.351^{\pm 0.043}$ & $164.8^{\pm 1.7}$ & $1.076^{\pm 0.014}$&$6.59^{\pm 0.26}$&$0.33^{\pm 0.11}$&$4.2^{\pm 10.2}$&$----$&$----$&$----$& $2.936^{\pm 0.024}$ & 1.57(42) \\ 
21 & Z-05-02 & 55691.5 & $0.094^{\pm 0.021}$ & $0.613^{\pm 0.059}$ & $255.9^{\pm 2.9}$ & $1.218^{\pm 0.023}$&$6.36^{\pm 0.37}$&$0.32^{\pm 0.09}$&$47.6^{\pm 8.3}$&$----$&$----$&$----$& $2.008^{\pm 0.018}$ & 1.46(42) \\
22 & Z-05-03 & 55693.0 & $0.063^{\pm 0.003}$ & $0.579^{\pm 0.049}$ & $302.1^{\pm 3.1}$&  $1.523^{\pm 0.027}$&$6.43^{\pm 0.42}$&$0.41^{\pm 0.07}$&$74.6^{\pm 9.2}$&$----$&$----$&$----$& $1.798^{\pm 0.012}$ & 1.21(42) \\
23 & Z-06-00 & 55694.0 & $0.046^{\pm 0.006}$ & $0.593^{\pm 0.029}$ & $332.2^{\pm 3.8}$&  $1.602^{\pm 0.048}$&$6.71^{\pm 0.48}$&$0.31^{\pm 0.25}$&$36.4^{\pm 6.5}$&$----$&$----$&$----$& $1.452^{\pm 0.009}$ & 1.59(42) \\
24 & Z-06-01 & 55695.4 & $0.031^{\pm 0.008}$ & $0.794^{\pm 0.033}$ & $344.2^{\pm 4.2}$&  $1.672^{\pm 0.042}$&$6.41^{\pm 0.34}$&$0.73^{\pm 0.27}$&$ 1.9^{\pm 9.8}$&$----$&$----$&$----$& $1.186^{\pm 0.007}$ & 1.16(42) \\ 
25 & Z-06-02 & 55697.1 & $0.012^{\pm 0.009}$ & $0.596^{\pm 0.048}$ & $375.9^{\pm 4.4}$&  $1.836^{\pm 0.032}$&$6.57^{\pm 0.28}$&$0.31^{\pm 0.14}$&$55.8^{\pm 6.8}$&$----$&$----$&$----$& $0.739^{\pm 0.005}$ & 0.81(42) \\ 
26 & Z-06-03 & 55698.5 & $0.016^{\pm 0.004}$ & $0.584^{\pm 0.054}$ & $438.1^{\pm 4.9}$&  $2.037^{\pm 0.043}$&$6.61^{\pm 0.19}$&$0.35^{\pm 0.16}$&$31.5^{\pm 5.2}$&$----$&$----$&$----$& $0.614^{\pm 0.003}$ & 1.01(42) \\ 
27 & Z-06-04 & 55700.2 & $0.017^{\pm 0.005}$ & $0.455^{\pm 0.049}$ & $446.5^{\pm 7.8}$&  $2.548^{\pm 0.055}$&$6.68^{\pm 0.16}$&$0.38^{\pm 0.22}$&$24.1^{\pm 4.6}$&$----$&$----$&$----$& $0.382^{\pm 0.002}$ & 1.02(42) \\

\hline
\end{tabular}
\vskip 0.4cm
\noindent{Here, Z=96425-01 is the initial part of observation IDs. Footnotes are the same as Appendix Table I.}

\end{table}

\clearpage

\begin{table}
\vskip -2.0cm
\addtolength{\tabcolsep}{-2.0pt}
\scriptsize                                                                                
	\centering 
        \centering{\large \bf Appendix Table III} 
	\caption{TCAF Model Fitted Mass and Normalization Values during 2010 outburst}
	\begin{tabular}{lcccccc} 
	\hline\noalign{\smallskip}
  Obs. & Id      &   Day         &  Normalization       &  Mass        & Mass& \\ 
       &         &  (MJD)        &                      & $(M_\odot)$  & $(M_\odot)$ \\ 
\hline                                                                              
 1& 95368-01-01-00 &  55417.3 &   $17.25^{\pm 1.12}$  &   $11.32^{\pm 0.44}$ &   $11.09^{\pm 0.46}$ \\
 2& 95360-14-01-00 &  55418.4 &   $15.93^{\pm 1.08}$  &   $11.59^{\pm 0.52}$ &   $12.16^{\pm 0.61}$ \\
 3& 95360-14-02-01 &  55419.1 &   $12.55^{\pm 0.91}$  &   $11.08^{\pm 0.48}$ &   $10.11^{\pm 0.42}$ \\
 4& 95360-14-02-00 &  55420.3 &   $15.04^{\pm 0.97}$  &   $11.53^{\pm 0.48}$ &   $11.93^{\pm 0.32}$ \\
 5& 95360-14-02-03 &  55422.0 &   $20.01^{\pm 1.35}$  &   $11.68^{\pm 0.56}$ &   $12.12^{\pm 0.47}$ \\
 6& 95360-14-03-01 &  55424.1 &   $15.37^{\pm 1.03}$  &   $11.35^{\pm 0.65}$ &   $10.12^{\pm 0.43}$ \\
 7& 95360-14-04-00 &  55425.2 &   $15.37^{\pm 1.03}$  &   $12.21^{\pm 0.52}$ &   $12.34^{\pm 0.57}$ \\ 
 8& 95360-14-05-00 &  55428.1 &   $19.21^{\pm 1.13}$  &   $11.75^{\pm 0.32}$ &   $11.91^{\pm 0.69}$ \\
 9& 95360-14-07-00 &  55431.1 &   $17.23^{\pm 1.11}$  &   $11.32^{\pm 0.42}$ &   $11.76^{\pm 0.59}$ \\
10& 95360-14-06-01 &  55433.1 &   $27.36^{\pm 1.32}$  &   $11.81^{\pm 0.38}$ &   $12.15^{\pm 0.39}$ \\
11& 95360-14-10-00 &  55438.2 &   $22.52^{\pm 1.22}$  &   $11.91^{\pm 0.52}$ &   $12.27^{\pm 0.42}$ \\
12& 95360-14-11-00 &  55439.1 &   $25.31^{\pm 1.28}$  &   $12.06^{\pm 0.66}$ &   $12.07^{\pm 0.55}$ \\
13& 95360-14-13-00 &  55442.9 &   $23.64^{\pm 1.11}$  &   $11.71^{\pm 0.58}$ &   $10.29^{\pm 0.32}$ \\
14& 95360-14-15-00 &  55444.6 &   $14.96^{\pm 0.98}$  &   $12.42^{\pm 0.67}$ &   $10.41^{\pm 0.25}$ \\
15& 95360-14-18-00 &  55447.8 &   $13.17^{\pm 0.92}$  &   $11.59^{\pm 0.46}$ &   $11.44^{\pm 0.29}$ \\
16& 95360-14-19-00 &  55448.8 &   $10.63^{\pm 0.79}$  &   $12.09^{\pm 0.71}$ &   $11.25^{\pm 0.29}$ \\
17& 95360-14-20-00 &  55450.3 &   $13.99^{\pm 1.03}$  &   $12.42^{\pm 0.69}$ &   $10.07^{\pm 0.36}$ \\
18& 95360-14-20-01 &  55451.2 &   $12.83^{\pm 0.78}$  &   $10.96^{\pm 0.36}$ &   $ 9.99^{\pm 0.22}$ \\
19& 95360-14-21-01 &  55453.7 &   $14.75^{\pm 1.05}$  &   $11.35^{\pm 0.48}$ &   $10.57^{\pm 0.26}$ \\
20& 95360-14-22-01 &  55455.4 &   $13.91^{\pm 1.15}$  &   $11.82^{\pm 0.62}$ &   $ 9.63^{\pm 0.28}$ \\
21& 95360-14-23-01 &  55457.1 &   $12.13^{\pm 1.09}$  &   $11.24^{\pm 0.56}$ &   $11.01^{\pm 0.51}$ \\
22& 95360-14-24-01 &  55459.7 &   $11.77^{\pm 1.07}$  &   $11.41^{\pm 0.51}$ &   $10.79^{\pm 0.36}$ \\
23& 95360-14-25-01 &  55462.6 &   $ 9.79^{\pm 0.85}$  &   $ 9.51^{\pm 0.33}$ &   $10.11^{\pm 0.33}$ \\
24& 95360-14-28-00 &  55465.1 &   $11.61^{\pm 0.88}$  &   $ 9.74^{\pm 0.42}$ &   $11.61^{\pm 0.49}$ \\
25& 95360-14-26-02 &  55467.5 &   $10.31^{\pm 0.75}$  &   $11.37^{\pm 0.43}$ &   $10.21^{\pm 0.52}$ \\
26& 95360-14-28-01 &  55469.0 &   $10.59^{\pm 0.72}$  &   $11.35^{\pm 0.52}$ &   $11.41^{\pm 0.44}$ \\
\hline\noalign{\smallskip}
\end{tabular}
\vskip 0.4cm
\noindent{In Columns 4 \& 5 TCAF fitted model normalization and mass of the black hole are shown during}
\noindent{2010 outburst when all parameters during fitting are kept free, and in Column 6 model fitted mass}
\noindent{is shown when normalization is frozen to $15.55$.}

\end{table}

\begin{table}
\vskip 0.1cm
\addtolength{\tabcolsep}{-2.0pt}
\scriptsize

	\centering
        \centering{\large \bf Appendix Table IV}
	\caption{TCAF Model Fitted Mass and Normalization Values during 2011 outburst}
	\begin{tabular}{lcccccc} 
	\hline
  Obs. & Id      &   Day         &  Normalization       &  Mass        & Mass& \\ 
       &         &  (MJD)        &                      & $(M_\odot)$  & $(M_\odot)$ \\ 

\hline                                                                              
 1 & 96425-01-01-00 & 55663.7 & $12.91^{\pm 0.86}$ & $10.12^{\pm 0.35}$ & $ 9.65^{\pm 0.31}$ \\
 2 & 96425-01-01-01 & 55665.9 & $10.81^{\pm 0.85}$ & $12.27^{\pm 0.56}$ & $ 9.66^{\pm 0.35}$ \\
 3 & 96425-01-02-00 & 55667.7 & $13.54^{\pm 0.95}$ & $12.02^{\pm 0.64}$ & $ 9.91^{\pm 0.44}$ \\
 4 & 96425-01-02-04 & 55668.5 & $13.02^{\pm 0.81}$ & $12.37^{\pm 0.53}$ & $ 9.25^{\pm 0.45}$ \\
 5 & 96425-01-02-01 & 55667.0 & $12.86^{\pm 0.76}$ & $12.42^{\pm 0.47}$ & $12.12^{\pm 0.32}$ \\
 6 & 96425-01-02-02 & 55670.7 & $14.17^{\pm 0.88}$ & $12.52^{\pm 0.61}$ & $11.04^{\pm 0.33}$ \\
 7 & 96425-01-02-05 & 55671.5 & $15.19^{\pm 1.07}$ & $12.71^{\pm 0.53}$ & $10.78^{\pm 0.31}$ \\ 
 8 & 96425-01-02-03 & 55672.9 & $13.14^{\pm 1.11}$ & $11.41^{\pm 0.38}$ & $12.86^{\pm 0.52}$ \\
 9 & 96425-01-03-00 & 55674.0 & $12.31^{\pm 1.01}$ & $12.12^{\pm 0.59}$ & $10.29^{\pm 0.48}$ \\
10 & 96425-01-03-05 & 55674.3 & $13.71^{\pm 1.07}$ & $12.51^{\pm 0.55}$ & $ 9.91^{\pm 0.36}$ \\
11 & 96425-01-03-01 & 55675.1 & $13.19^{\pm 0.87}$ & $ 9.43^{\pm 0.28}$ & $ 9.92^{\pm 0.34}$ \\
12 & 96425-01-03-02 & 55676.4 & $16.18^{\pm 0.98}$ & $12.52^{\pm 0.52}$ & $10.22^{\pm 0.41}$ \\
13 & 96425-01-03-03 & 55678.1 & $13.29^{\pm 1.04}$ & $11.28^{\pm 0.49}$ & $12.33^{\pm 0.52}$ \\
14 & 96425-01-03-04 & 55679.3 & $13.27^{\pm 1.07}$ & $10.88^{\pm 0.42}$ & $12.41^{\pm 0.55}$ \\
15 & 96425-01-04-02 & 55680.2 & $12.27^{\pm 0.77}$ & $ 9.89^{\pm 0.35}$ & $10.69^{\pm 0.49}$ \\
16 & 96425-01-04-00 & 55681.2 & $11.21^{\pm 0.88}$ & $12.29^{\pm 0.62}$ & $12.42^{\pm 0.54}$ \\
17 & 96425-01-04-03 & 55682.2 & $11.44^{\pm 1.01}$ & $11.49^{\pm 0.48}$ & $11.79^{\pm 0.49}$ \\
18 & 96425-01-04-01 & 55684.7 & $14.27^{\pm 1.05}$ & $11.08^{\pm 0.38}$ & $11.49^{\pm 0.48}$ \\
19 & 96425-01-05-00 & 55687.6 & $11.51^{\pm 1.09}$ & $10.29^{\pm 0.33}$ & $10.87^{\pm 0.37}$ \\
20 & 96425-01-05-01 & 55690.1 & $11.75^{\pm 1.12}$ & $12.34^{\pm 0.56}$ & $11.43^{\pm 0.48}$ \\
21 & 96425-01-05-02 & 55691.5 & $10.71^{\pm 0.75}$ & $12.17^{\pm 0.53}$ & $ 9.35^{\pm 0.41}$ \\
22 & 96425-01-05-03 & 55693.0 & $11.97^{\pm 0.82}$ & $11.93^{\pm 0.56}$ & $10.49^{\pm 0.55}$ \\
23 & 96425-01-06-00 & 55694.0 & $11.11^{\pm 1.05}$ & $11.95^{\pm 0.62}$ & $ 9.73^{\pm 0.42}$ \\
24 & 96425-01-06-01 & 55695.4 & $10.12^{\pm 0.75}$ & $12.45^{\pm 0.71}$ & $10.92^{\pm 0.45}$ \\
25 & 96425-01-06-02 & 55697.1 & $11.11^{\pm 1.06}$ & $10.71^{\pm 0.28}$ & $11.68^{\pm 0.48}$ \\
26 & 96425-01-06-03 & 55698.5 & $10.01^{\pm 0.89}$ & $ 9.71^{\pm 0.24}$ & $11.27^{\pm 0.37}$ \\
27 & 96425-01-06-04 & 55700.2 & $12.21^{\pm 0.97}$ & $10.51^{\pm 0.49}$ & $11.28^{\pm 0.34}$ \\
\hline
\end{tabular}
\vskip 0.4cm

\noindent{Footnotes are same as Appendix Table III}

\end{table}

\end{document}